\documentclass[useAMS,usenatbib]{mn2e}
\usepackage{psfig}
\usepackage{amssymb}
\usepackage{color}
\usepackage{enumerate}
\usepackage{rotating}
\usepackage{amsmath}
\usepackage{mathrsfs}
\usepackage{ragged2e}
\usepackage{hyperref}

\title[The High-Mass VDF from BOSS]
{A Direct Measurement of the High-Mass End of the Velocity Dispersion Function at $z\sim0.55$ from SDSS-III/BOSS \vspace{-0.5cm}}

\author[Montero-Dorta et al.]{
\parbox[t]{\textwidth}{
Antonio D. Montero-Dorta$^{1}$\thanks{E-mail: amontero@astro.utah.edu}, Adam S. Bolton$^{2,1}$ \& Yiping Shu$^{3}$}
\vspace*{6pt} \\ 
$^1$ Department of Physics and Astronomy, The University of Utah, 115
South 1400 East, Salt Lake City, UT 84112, USA \\
$^2$ National Optical Astronomy Observatory (NOAO), 950 North Cherry Ave., Tucson, AZ 85719, USA \\
$^3$ National Astronomical Observatories, Chinese Academy of Sciences, 20A Datun Road, Chaoyang District, Beijing 100012, China
\vspace{-0.4cm} 
}

\date{Accepted ---. Received ---;in original form --- \vspace{-0.3cm}}

\def\simlt{\lower.5ex\hbox{$\; \buildrel < \over \sim \;$}}
\def\simgt{\lower.5ex\hbox{$\; \buildrel > \over \sim \;$}}
\usepackage{graphicx}
\usepackage{rotating}

\definecolor{red}{rgb}{1,0,0}

\begin{document}

\bibliographystyle{mnras}

\maketitle

\begin{abstract}

We report the first direct spectroscopic measurement of the velocity dispersion function (VDF) 
for the high-mass red sequence (RS) galaxy population at redshift $z\sim0.55$. We achieve 
high precision by using a sample of 600,000 massive galaxies with spectra from the Baryon 
Oscillation Spectroscopic Survey (BOSS) of the third Sloan Digital Sky Survey (SDSS-III), covering 
stellar masses $M_*\gtrsim10^{11}~M_{\odot}$. We determine the VDF by projecting the 
joint probability-density function (PDF) of luminosity $L$ and velocity dispersion $\sigma$, i.e. $p(L,\sigma)$, 
defined by our previous measurements of the RS luminosity function and $L-\sigma$ relation for this sample.
These measurements were corrected from red--blue galaxy population 
confusion, photometric blurring, incompleteness and selection effects within a forward-modeling 
framework that furthermore correctly accommodates the low spectroscopic signal-to-noise ratio 
of individual BOSS spectra. Our $z\sim0.55$ RS VDF is in overall agreement with the $z\sim0$ early-type 
galaxy (ETG) VDF at $\log_{10}\sigma\gtrsim2.47$, however the number density
of $z=0.55$ RS galaxies that we report is larger than that of $z=0$ ETG galaxies at $2.35\gtrsim\log_{10}\sigma\gtrsim 2.47$. 
The extrapolation of an intermediate-mass L-$\sigma$ relation towards the high-mass end in previous low-z works
may be responsible for this disagreement. Evolutionary interpretation of this comparison 
is also subject to differences in the way the respective samples are selected; these differences can be mitigated in future 
work by analyzing $z=0$ SDSS data using the same framework presented in this paper. We also provide the sample 
PDF for the RS population (i.e. uncorrected for incompleteness), which is a key ingredient for gravitational lensing analyses 
using BOSS.

\end{abstract}


\section{Introduction}
\label{sec:intro}

The Baryon Oscillation Spectroscopic Survey (BOSS, \citealt{Schlegel2009_w, Dawson2013}) of 
the SDSS-III \citep{Eisenstein2011} is the largest dark-energy (DE) experiment to date. With the aim of 
understanding the nature of the DE that drives the present-day accelerated expansion of the Universe, 
BOSS has collected a massive sample of $\sim1.5$ million galaxies, most of them Luminous Red Galaxies (LRGs). This sample
is used to map the large-scale structure (LSS) of the Universe with significant accuracy, in order to measure the Baryon Acoustic Oscillations (BAO) 
from which cosmological constraints are derived (see, e.g., \citealt{Anderson2014}, \citealt{Aubourg2015}). In addition to its cosmological value, a 
sample of such size can be used to study the evolution of the properties of massive galaxies (see, e.g., \citealt{Shu2012,Tojeiro2012,
Maraston2013,Beifiori2014,MonteroDorta2016A,MonteroDorta2016B,Bernardi2016}).

BOSS provides unprecedented statistical coverage of massive galaxies by surveying a huge volume of the Universe; 
however, this coverage is achieved at the expense of non-trivial colour--magnitude selection cuts and relatively low 
spectroscopic signal-to-noise (S/N), which together present significant challenges to the application of the sample to the study of galaxy evolution.
The majority of DE surveys work in this low-S/N, large-N regime. Low S/N spectra and higher 
redshift implies larger photometric errors that distort the distribution of photometric observables and 
hinder our ability to distinguish between the intrinsically red and blue galaxy populations. In \cite{MonteroDorta2016A} (hereafter, MD16A) 
and \cite{MonteroDorta2016B} (MD16B), we present a forward-modeling framework based on 
Bayesian inference that is suitable for the study of the main properties of BOSS galaxies. We have 
successfully applied this method to the deconvolution of both photometric 
(MD16A) and spectroscopic (MD16B) properties of the high-mass red sequence (RS) galaxy population at $z\sim0.55$, from 
the effects of selection, photometric errors and low S/N data.

The aforementioned methodology allows us to characterize incompleteness in the BOSS CMASS sample, 
which provides photometric and spectroscopic information for more than 1 million massive galaxies
in the redshift range $0.4<z<0.7$ (CMASS stands for ``Constant MASS''). More importantly, it allows us to 
model the latent parameters of the high-mass RS galaxy population. 
We have shown that the RS galaxy population forms an extremely 
compact distribution in the optical colour-colour plane (consistent with a delta function for a given 
magnitude and redshift bin, see MD16A). In addition, the high-mass RS appears to evolve passively (or very close), 
as inferred from the evolution of the RS luminosity function (RS LF, see MD16A). In MD16B, we also show
that scaling relations, i.e. the L-$\sigma$ relation, for the high mass RS are very different from what it was measured from their 
intermediate-mass counterparts, showing a significantly steeper slope and smaller scatter. 
In the present work, we complement this picture by measuring the RS velocity dispersion function (VDF) 
at $z\sim0.55$. The VDF is defined as the number density of galaxies per logarithmic decade of the stellar velocity dispersion $\sigma$. 

The stellar velocity dispersion is a fundamental quantity in the study of early-type galaxies (ETGs), since it is tightly 
connected to their total mass and carries information about the physical 
processes that shape their evolution. The stellar velocity dispersion is one of the 3 physical properties 
that constitute the so-called {\it{fundamental plane}} of ETGs, along with the effective radius and the surface brightness
(\citealt{Djorgovski1987}, \citealt{Dressler1987}). As a statistical property of the galaxy 
population, the VDF has the advantage, as compared to the LF or the stellar mass function (SMF), in that it is completely independent of 
stellar population synthesis models (i.e., {\it{k+e corrections}}). The lack of adequate high-z spectroscopic samples
has hindered, however, the use of the VDF as a galaxy-evolution probe.

At low redshifts, the VDF has been directly measured 
using massive spectroscopic surveys like the SDSS-I (\citealt{Sheth2003, Mitchell2005, Choi2007, Chae2010}), which 
mostly cover the intermediate-mass range. At higher redshifts, direct kinematic measurements  
have not been reported, due to a combination of selection effects, low S/N spectra and/or small footprint. Estimates for the high-z VDF have only been inferred indirectly
from other observational proxies such as strong-lensing statistics \citep{Chae2010} or using photometric predictions based on low-z relations \citep{Bezanson2011}.
The predictions from \cite{Bezanson2011} suggest that the VDF at $z=0.5$ may be very similar to the VDF at $z=0$. At higher redshifts, however,
their results suggest a decrease in the number density of low-$\sigma$ galaxies, while the number density of high-$\sigma$
galaxies remains constant or increases. These predictions are in some tension with the strong-lensing-constrained results from 
\cite{Chae2010}, who claims that the VDF remains fairly constant at the low-dispersion end ($\log_{10} \sigma < 2.3$), but the number of higher-$\sigma$
galaxies decreases with redshift.

The direct kinematic measurement that we provide in this paper at $z=0.55$ is therefore the first 
measurement of this kind at $z\gtrsim0.2$. In addition to its value as a means of characterizing the
galaxy population and its evolution, the VDF can be used to study the connexion between 
galaxies and haloes. Instead of the LF or the SMF, the VDF can be used as an input for halo abundance matching models 
(HAMs: e.g., \citealt{Vale2004}; \citealt{Trujillo2011}) or halo occupation distributions (HOD: e.g., \citealt{Berlind2002}; \citealt{Zehavi2005}), 
in combination with N-body cosmological simulations (see an example in \citealt{Chae2010}). In fact, \cite{Wake2012}  
show that the velocity dispersion is more closely related to the clustering amplitude of galaxies than either stellar or dynamical mass. 

The VDF is also a key ingredient for statistical computations of the incidence of strong gravitational lensing
(see, e.g., \citealt{Wambsganss1998,Bartelmann2010,Treu2010}, for a comprehensive reviews on 
strong gravitational lensing). The VDF translates directly into the integrated cross section for strong lensing to a given redshift within a given cosmology (see, e.g. \citealt{Turner1984, Mitchell2005}). 
This cross section can in turn be used to predict the incidence of strong lenses for a given source population, or to infer 
cosmological parameters or lens mass-distribution parameters from the statistics of an observational strong-lens sample (see, e.g.,
 \citealt{Chae2002, Chae2003, Browne2003, Mitchell2005}, using the Cosmic Lens All-Sky Survey, CLASS, \citealt{Myers1995,Browne2003}). 
The VDF can also be combined with observations of weak gravitational lensing (see \citealt{Bartelmann2001,Refregier2003,Hoekstra2008}
for comprehensive reviews on weak gravitational lensing)
on larger projected length scales to constrain statistical models for the occupation of galaxies within dark-matter 
halos and the cosmological parameters (\citealt{Sheldon2004, Mandelbaum2013,Miyatake2015,More2015}).

This paper is organized as follows. The data used in our analysis 
is briefly described in Section~\ref{sec:data}. 
In section~\ref{sec:VDF_method}, we describe the method employed to compute 
the VDF and we provide a brief summary of our previous results on the RS LF (MD16A)
and L-$\sigma$ relation (MD16B) (\ref{sec:ingredients}). In 
Section~\ref{sec:results}, we present the main results of our analysis, including 
the $z=0.55$ RS VDF (\ref{sec:VDF}), a comparison with previous VDF measurements (\ref{sec:evolution})
and the sample PDF for the RS (\ref{sec:pdf}). Finally, 
in Section~\ref{sec:conclusions} we summarize the main conclusions of our work and 
discuss some applications of our results. Throughout this paper we adopt a cosmology 
with $\Omega_M=0.274$,  $\Omega_\Lambda=0.726$ and $H_0 = 100h$ km s$^{-1}$ Mpc$^{-1}$ with $h=0.70$ 
(WMAP7, \citealt{Komatsu2011}), and use AB magnitudes \citep{OkeGunn1983}.

\section{The data} 
\label{sec:data}

In this work, we combine LF results from MD16A and L-$\sigma$ relation results from 
MD16B to compute the high-mass VDF at $z\sim0.55$. These previous results are obtained 
using both spectroscopic and photometric data from the Tenth Data Release of the SDSS 
(DR10, \citealt{Ahn2014}). The DR10 is also the third data release of the SDSS-III program and the second 
release that includes BOSS data. 

The spectroscopic DR10 BOSS sample contains a total of $927,844$ galaxy 
spectra and $535,995$ quasar spectra. The baseline 
imaging sample for the DR10 is the final SDSS imaging data set, which contains, not only the new SDSS-III imaging, but also the previous SDSS-I and II imaging data
(this imagining data set was released as part of the DR8, see \citealt{Aihara2011}). The imagining programs provide five-band {\it{ugriz}} 
imaging over 7600 sq deg in the Northern Galactic Hemisphere and $\sim$ 3100 sq deg in the Southern Galactic 
Hemisphere. The $50\%$ completeness limit for detection of point sources corresponds to a 
typical magnitude of $r = 22.5$. Comprehensive information about technical aspects of the SDSS survey
can be found in the following papers:
\cite{Fukugita1996} describes the SDSS {\it{ugriz}} photometric system; 
\cite{Gunn1998} and \cite{Gunn2006} describe the SDSS camera and the SDSS telescope, respectively;
\cite{Smee2013} provides detailed information about the SDSS/BOSS spectrographs. 

As in both MD16A and MD16B, we restrict our analysis to the CMASS  
spectroscopic sample of the BOSS spectroscopic catalog. The CMASS sample is mostly 
comprised by LRGs and covers a nominal redshift range $0.43<z<0.70$ (the 
mean redshift is $0.532$). The stellar masses for the red population, as measured by \cite{Maraston2013},  
are $M_* \gtrsim 10^{11} M_{\odot}$, peaking at $M_* \simeq 10^{11.3} M_{\odot}$, assuming a Kroupa 
initial stellar mass function \citep{Kroupa2001}. The total number of unique CMASS galaxies with a good redshift estimate and with \texttt{model} and \texttt{cmodel} apparent magnitudes 
and photometric errors in all g,r and i bands in the catalog is 549,005. For more information 
on the BOSS selection refer to \cite{Eisenstein2011}, \cite{Dawson2013} and \cite{Reid2016}. A 
brief summary is also provided in MD16A and MD16B. For a complete discussion on the selection effects affecting the CMASS data, see MD16A.

One of the key ingredients in the L-$\sigma$ relation study presented in MD16B is the likelihood function of the central stellar velocity dispersion for each galaxy. 
The method for determining this likelihood function is described in detail in \cite{Shu2012}\footnote{See \cite{Thomas2013}
for a complete discussion on the different stellar velocity dispersion estimates available in BOSS.}. In essence, for every galaxy in the BOSS samples,
 the line-of-sight stellar velocity dispersion within the central circular region of radius $1$ arcsec
is measured spectroscopically by fitting a linear combination of broadened stellar {\it{eigenspectra}} to the observed galaxy 
spectrum (note that the typical seeing for BOSS is $1.5$ arcsec). Subsequently, the $\chi^2$ of the fit as a function of the trial velocity dispersion is
converted into the likelihood function of velocity dispersion given the observational data.

As discussed in MD16B, due to the low S/N of the BOSS spectra, the velocity-dispersion likelihood functions are not sufficiently 
Gaussian to permit unbiased point estimates and simple error bars for the stellar velocity dispersion on a galaxy-by-galaxy basis.
As a reference, however, the distribution of the best-fit central velocity dispersion for the CMASS sample (as inferred from the likelihood functions) is centered 
around $\log_{10} \sigma \simeq 2.35$, or 220 km/s, and $\Delta \log_{10} \sigma \simeq 0.06$ dex, or 30 km/s. More than two-thirds of 
the subsample have velocity dispersions between $120$ km/s and $300$ km/s with $20-50$ km/s uncertainties. 

\begin{figure}
\begin{center}
\includegraphics[scale=0.5]{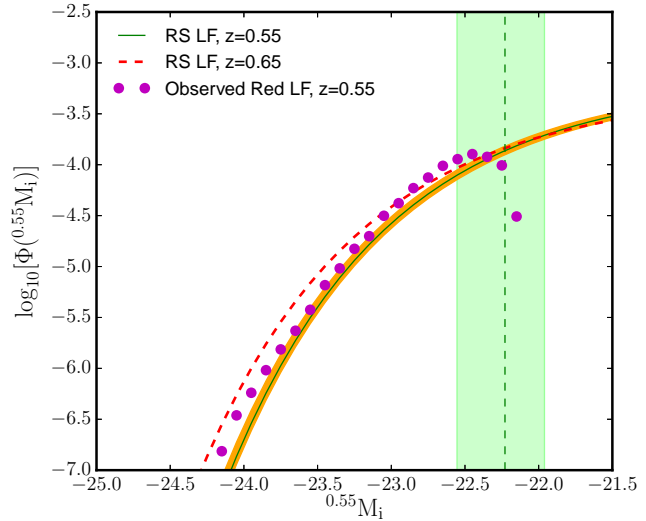}
\caption{The $^{0.55}i$-band RS LF derived from the BOSS CMASS sample as 
presented in MD16A. The solid line shows our fiducial model at $z=0.55$.  
The shaded regions around this line represent the uncertainty in the determination of the LF, where both 
the statistical and the systematic errors are taken into account. The dashed line shows, for reference, the 
RS LF at $z=0.65$ (no evolution correction has been applied). The vertical line indicates the 50\%-completeness limit of the sample; the shaded region
around this line shows the 2\% and 98\% completeness limits, respectively. An ``observed"
RS LF is provided in dots in order to illustrate the effect of the red--blue deconvolution and 
the completeness correction. This observed red LF has been computed from an 
object-by-object standpoint by applying a simple colour cut $(g-i) > 2.35$ to isolate 
the red galaxy population in observed space.} 
\label{fig:lf}
\end{center}
\end{figure}

\begin{figure}
\begin{center}
\includegraphics[scale=0.53]{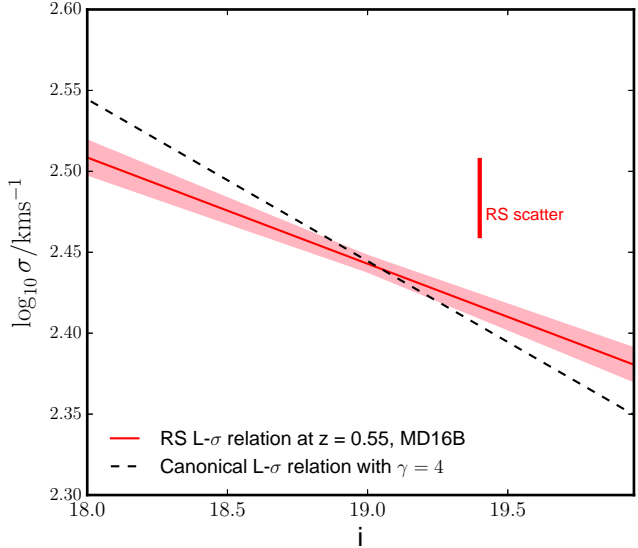}
\caption{The $<\log_{10} \sigma> $ - i-band apparent magnitude relation at $z=0.55$ derived from the BOSS CMASS sample
for the RS (red line) as presented in MD16B. Uncertainties
are represented by shaded regions. The dashed line shows, for reference, an ad hoc RS relation
where the slope $\gamma$ is assumed to be equal to the canonical value of 4 (the 
shape and zero-point of this relation are kept unchanged). The vertical line
represent the intrinsic scatter in $\log_{10} \sigma$, measured for the RS.} 
\label{fig:lsigma}
\end{center}
\end{figure}

\section{Computation of the VDF} 
\label{sec:VDF_method}

The VDF, which we will denote by $\Phi(\sigma)$, is defined as the (intrinsic) number density of objects 
per unit stellar velocity dispersion. Namely:
 
\begin{eqnarray} \displaystyle
dN = \Phi(\sigma)~d\sigma dV
\label{eq:vdf}
\end{eqnarray}

\noindent where $dN$ is the total number of objects within the range $[\sigma,\sigma+d\sigma]$ and 
comoving volume element $d V$. 

Since the BOSS CMASS sample is selected in colour-magnitude space with a non-trivial 
selection function, we cannot work exclusively in velocity-dispersion space if we wish to 
derive the intrinsic $\Phi(\sigma)$ corrected for incompleteness. Instead, we need
to work in $L-\sigma$ space and make used of the previously measured 
LF (MD16A) and $L-\sigma$ relation (MD16B). 

In statistics terminology, $\Phi(\sigma)$ is closely related to the marginal 
probability distribution for $\sigma$, $p(\sigma)$ . The only difference is that $\Phi(\sigma)$ is 
normalized per unit volume, while $p(\sigma)$ is normalized by number. The marginal probability distribution
$p(\sigma)$ can be obtained by projecting the joint PDF of $L$ and $\sigma$, i.e., $p(L,\sigma)$. Since 
$p(L,\sigma) = p(\sigma|L) \times p(L)$, $p(\sigma)$ can be obtained by computing the 
following integral over $L$:

\begin{eqnarray} \displaystyle
p(\sigma) = \int{ dL p(L)~p(\sigma|L)}
\label{eq:vdf_prob}
\end{eqnarray}

\noindent where $p(L)$ is the marginal probability distribution for L, $p(L)$, and $p(\sigma|L)$ is the 
conditional probability distribution of $\sigma$ given $L$. It can be easily shown that 
$\Phi(\sigma)$ can be derived in a similar way, namely:

\begin{eqnarray} \displaystyle
\Phi(\sigma) = \int{ dL \phi(L)~p(\sigma|L)}
\label{eq:vdf}
\end{eqnarray}

\noindent where $\phi(L)$ is the LF, i.e the marginal probability distribution normalized per unit 
volume. Since $p(\sigma|L)$ encodes the $L-\sigma$ relation, we have arrived at an expression entirely in terms of 
quantities that we have previously measured.

In practice, it is convenient to work in absolute magnitude M - $\log_{10} \sigma$ space. In this context, 
$\Phi(\log_{10} \sigma)$ can be obtained from $\phi(M)$ and $p(\log_{10} \sigma|M)$. 
The LF is often parametrized in magnitude space as a Schechter Function \citep{Schechter1976} of the form:

\begin{eqnarray} \displaystyle
\phi (M,\{\phi_*,M_*, \alpha\}) = 0.4 \log(10) \phi_* \left[10^{0.4(M-M_*)(\alpha+1)}\right] \times  \nonumber \\
exp\left(-10^{0.4(M-M_*)}\right) 
\label{eq:schechter}
\end{eqnarray}

\noindent where $\phi_*$,$M_*$, $\alpha$ are the 3 Schechter parameters that describe the normalization, characteristic 
magnitude and faint-end slope of the LF, respectively.  

Motivated by results from \cite{Bernardi2003b}, the conditional PDF $p(\log_{10} \sigma|M)$ can be approximated by a Gaussian 
PDF in $\log_{10} \sigma$ with mean $<\log_{10} \sigma>$ and intrinsic scatter $s$, of the form:

\begin{equation}
P (\log_{10} \sigma | \mathbf{M}) = \frac{1}{\sqrt{2 \pi} s} \exp\left[\frac{-(\log_{10} \sigma - <\log_{10} \sigma>)^2}{2 s^2}\right]
\label{eq:int_pdf}
\end{equation}

\noindent where the dependence on magnitude enters through $<\log_{10} \sigma>$. The mean velocity dispersion 
depends on the luminosity of the galaxy as $\sigma \propto L^{1/\gamma}$ (or $\log_{10} \sigma \propto [- 0.4/\gamma]~M$). We 
generically refer to this relation as the L-$\sigma$ relation \citep{Minkowski1962}. The special case where 
$\gamma = 4$ is called the Faber-Jackson relation \citep{FJR}.

\begin{figure*}
\begin{center}
\includegraphics[scale=0.70]{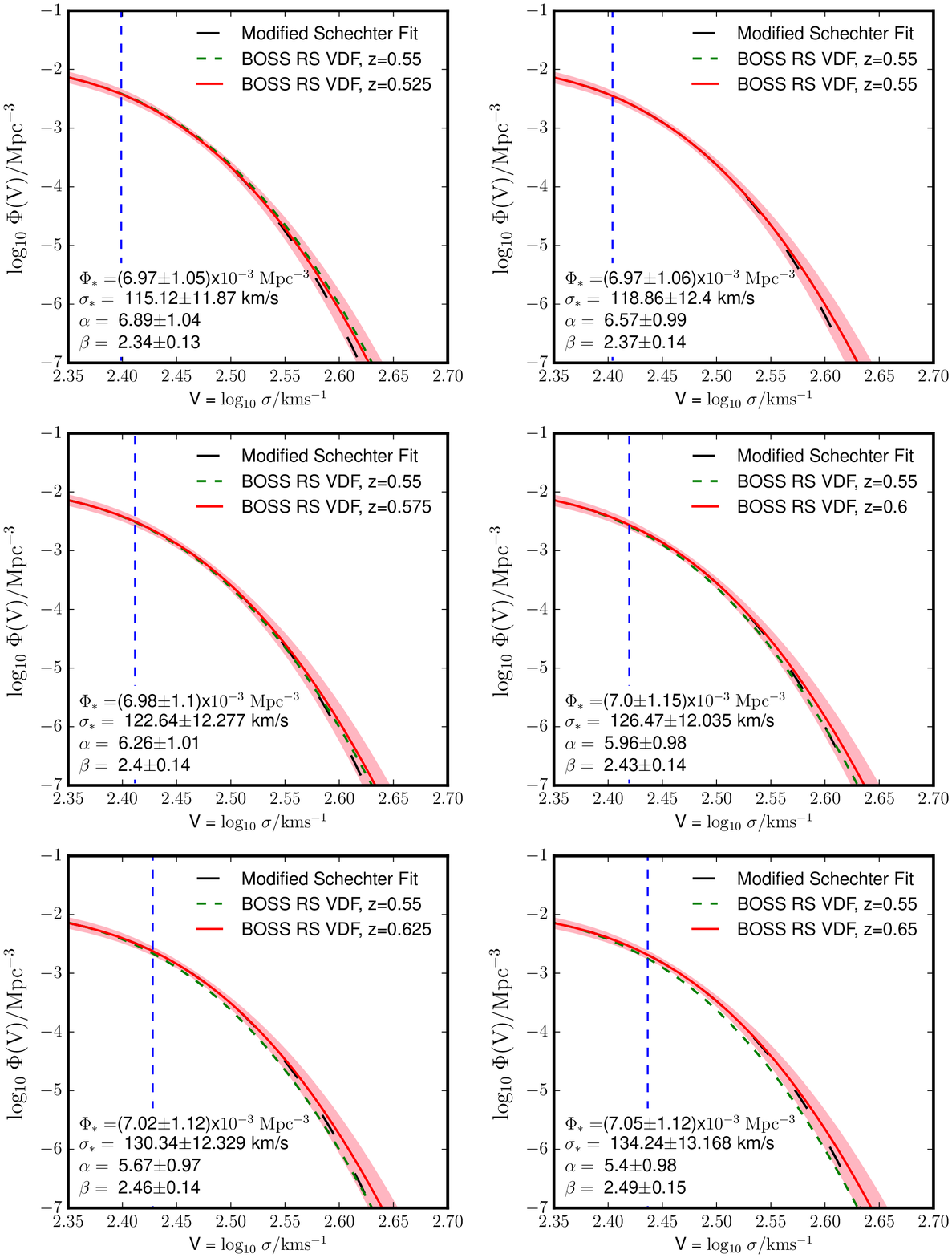}
\caption{The high-mass end of the RS VDF as measured from the BOSS
CMASS sample in six redshift slices around $z=0.55$. The solid line in each panel shows the RS VDF at the corresponding 
redshift, and the long-dashed line the best-fit modified Schechter Function (see text). Best-fit parameters are also provided. The short-dashed line provides the VDF 
at the reference redshift $z=0.55$. The vertical line indicates the $50\%$ completeness limit of the sample
in velocity dispersion, $\sigma_{50}$. Shaded regions indicate the 1-$\sigma$ statistical uncertainty. } 
\label{fig:vdf_boss}
\end{center}
\end{figure*}

\subsection{The high-mass RS at $z\sim0.55$ from BOSS: LF and L-$\sigma$ relation.} 
\label{sec:ingredients}

In MD16A, the intrinsic RS colour-colour-magnitude distribution in the CMASS sample is deconvolved from 
photometric errors, selection effects and red--blue galaxy population confusion using 
forward-modeling techniques, within a hierarchical Bayesian framework. This intrinsic distribution model forms the basis of 
the LF and L-$\sigma$ relation presented in MD16A and MD16B, respectively. 
Our modeling is based on a phenomenological approach, intended to 
describe the bimodality found in the colour-colour plane, thus we avoid making use of 
previous assumptions about ``blue" or ``red" galaxies or stellar population synthesis models. As a brief summary, the two components 
of the intrinsic model present the following general characteristics:
 
\begin{itemize}
	\item {\it{Red Sequence (RS)}}: The RS is so narrow that is consistent, within the errors, at fixed magnitude and for a narrow redshift slice, with a delta function in 
	the colour-colour plane (width $<0.05$ mag), with only a shallow colour-magnitude relation shifting the location of this point. 
	\item {\it{Blue Cloud (BC)}}: The BC is defined as a background distribution that contains {\it{everything not
	belonging to the RS}} and is well described by a more extended 2-D Gaussian in the colour-colour plane. The
	RS is superimposed upon the BC, since the latter extends through the red side of the colour-colour plane. Our BC is a spectroscopically 
	and photometrically heterogeneous population that can be clearly distinguished from the RS in intrinsic space, but 
	not necessarily restricted to a population of ``blue" or ``young" galaxies \footnote{In this sense, the construction of our RS/BC model differs  
	significantly from the way the ETG SDSS samples are selected, based on morphological properties (see \citealt{Bernardi2003a} as an example).}.
\end{itemize}

In this paper, we restrict our computation of the VDF to the RS, since the BC is severely incomplete in the 
CMASS sample (see MD06A). The first element needed in Equation~\ref{eq:vdf}, the RS LF, 
is derived in MD16A from the intrinsic RS colour-colour-magnitude distribution. 
The RS LF at $z=0.55$ in the $^{0.55}i$ band (the i band K-corrected to $z=0.55$) is presented in Figure~\ref{fig:lf}
(the $z=0.65$ RS LF is also shown for reference). The CMASS sample maps with unprecedented
precision the very-bright end of the RS; the vertical line in Figure~\ref{fig:lf} represents the $50\%$-completeness limit in the sample, i.e. $^{0.55}M_i \lesssim -22.25$ at $z=0.55$. 
An ``observed" red LF obtained from the CMASS sample\footnote{By ``observed''
here we mean ``not error-deconvolved", ``not intrinsic''. The ``observed" red LF has been computed from an 
object-by-object standpoint by applying a simple colour cut $(g-i) > 2.35$ (see MD16A) to isolate 
the red galaxy population in observed space.} is also shown in Figure~\ref{fig:lf},
in order to illustrate the effect of the photometric deconvolution on the determination of the RS LF.

The following linear relations to the Schechter parameters describe the redshift evolution of the RS LF between $0.525<z<0.63$:

\begin{eqnarray} \displaystyle
\Phi_*(z) = [(-0.189\pm0.372)~z + (0.834\pm0.216)] \times \\ 
\times10^{-3}~{\rm{Mpc}}^{-3}~{\rm{mag}}^{-1}\nonumber \\ 
^{0.55}M_*(z) = (-1.943\pm0.228)~z  + (- 20.658\pm0.132)
\label{eq:linear_fits2}
\end{eqnarray}	
	
\noindent Note that, although the CMASS expands a redshift range $0.4\lesssim z \lesssim 0.7$, in practice, as 
discussed in MD16A, the computation of the LF had to be restricted to a narrower redshift range, in order 
to minimize the effect of systematic errors. 

The second element needed to compute the VDF is the RS L-$\sigma$ relation, which enters Equation~\ref{eq:int_pdf}.
The RS L-$\sigma$ relation at $z\sim0.55$ is derived from the CMASS sample in MD16B using
the PDSO method (i.e., Photometric Deconvolution of Spectroscopic Observables). The PDSO
method allows us to extend the photometric deconvolution of the intrinsic RS/BC galaxy distributions 
to luminosity-velocity dispersion space. Our results indicate that the high-mass RS L-$\sigma$ relation at $z\sim0.55$ is significantly 
steeper than the canonical relation found at lower mass ranges (a slope of $\gamma \simeq 8$ instead of 4). 
The scatter in $\log_{10}\sigma$ is also at most half the values that have been measured for intermediate masses. 
This differences are relevant since previous direct measurements of the VDF (all of them at low redshift)
are based on intermediate-mass samples such as the SDSS. Our L-$\sigma$ relation results are illustrated in Figure~\ref{fig:lsigma}. 
We show, for convenience, the $<\log_{10} \sigma>$ - i (apparent magnitude) relation at $z=0.55$ for the RS, along with an 
ad hoc canonical relation with the same shape and zero-point as the former but a slope of $\gamma = 4$. 
It is important to emphasize again that the red--blue deconvolution is performed using both photometric and spectroscopic information.

The RS L-$\sigma$ relation from BOSS is parametrized in MD16B using the following expression:

\begin{equation}
<\log_{10} \sigma> = c_{1} + 2.5 + c_{2} (^{0.55}M_i + 23)
\label{eq:F-J relation}
\end{equation}

\noindent where $c_{1}$ corresponds to the zero point and $c_{2}$ to the slope. The redshift 
evolution of these parameters is found to be well-described by the following linear relations:

\begin{equation}
c_{1} (z) = 2.429\pm0.007 + (0.023\pm0.011)~z
\label{eq:m2}
\end{equation}

\begin{equation}
c_{2} (z) = - (0.033\pm0.012) - (0.029\pm0.021)~z
\label{eq:m2}
\end{equation}

\noindent and the scatter in $\log_{10} \sigma$ measured for the distribution shown in Equation~\ref{eq:int_pdf} is $s = 0.047\pm0.004$.

\section{Results} 
\label{sec:results}

\subsection{The high-mass end of the VDF at $z\sim0.55$} 
\label{sec:VDF}

In Figure~\ref{fig:vdf_boss}, we show the high-mass end of the RS VDF as measured from the 
CMASS sample in 6 different redshift bins from $z=0.525$ to $z=0.65$, where the bin size used is $\Delta z = \pm0.01$. 
The integration in Equation~\ref{eq:vdf} is performed across the entire magnitude range $[-\infty,+\infty]$. This 
computation is only valid within the CMASS luminosity/range. An extrapolation beyond this 
range would be break down because both the LF and the L-$\sigma$ relation are only 
complete at the high-mass end. Also, note that the 
L-$\sigma$ relation has a curved shape, so the same relation does not hold for lower-mass ranges. The vertical line
indicates the approximate $50\%$-$\sigma$-completeness threshold in the sample, $\sigma_{50}$. This threshold
was computed directly transforming the $50\%$ absolute-magnitude completeness limit (MD16A)
into a velocity dispersion limit using the L-$\sigma$ relation (MD16B). This 
is an approximation that works well because the scatter in the L-$\sigma$ relation is very small. 
Only results above the blue dashed line are fully reliable. 

The errors shown for the VDF (shaded regions in Figure~\ref{fig:vdf_boss}) correspond to the 
scatter measured with respect to the best-fit linear relation as a function of redshift for each number density considered. 
We have found that the uncertainty in the L-$\sigma$ relation is, by far, the largest 
contributor to the total error budget of our VDF measurement. This conclusion is
based on a Monte Carlo analysis where we measure the scatter that propagates into 
the VDF when only the uncertainties associated to either the LF or the L-$\sigma$ relation
are considered. This scatter is $\sim5$ larger when only the errors on the latter are taken into account
(and the LF is assumed to be noise-free). Note that the 
errors on these 2 statistical measurements are correlated, since the LF is used in the determination of the 
L-$\sigma$ relation.

\begin{figure}
\begin{center}
\includegraphics[scale=0.35]{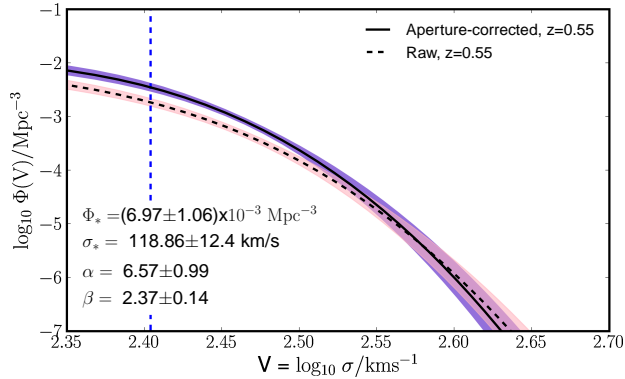}
\caption{Aperture-corrected vs. raw high-mass RS VDF at $z=0.55$. Shaded regions indicate the 1-$\sigma$ statistical uncertainty. 
The vertical line indicates the $50\%$ completeness limit of the sample
in velocity dispersion, $\sigma_{50}$.
Best-fit parameters for the modified Schechter Function are provided for the raw VDF.} 
\label{fig:vdf_nonac}
\end{center}
\end{figure}

The VDF is commonly parametrized since \cite{Sheth2003} using a modified Schechter function 
of the form:

\begin{eqnarray} \displaystyle
\Phi(\sigma) d \sigma =  \Phi_* \left(\frac{\sigma}{\sigma_*}\right)^{\alpha} \frac{\exp[(-\sigma / \sigma_*)^{\beta}]}{\Gamma(\alpha/\beta)} \beta \frac{d \sigma}{\sigma}
\label{eq:vdf_fit}
\end{eqnarray}

In each panel of Figure~\ref{fig:vdf_boss}, we also show the best-fit Schechter-like Function along with the best-fit parameters for 
the high-mass RS VDF. The modified Schechter function provides a good description of the 
the high-mass RS VDF, although it underestimates the extremely high-mass end (this is
also well-documented for the LF). Note that the Schechter parameters are highly covariant and nearly degenerate
at the high-mass end, so caution must be exercised when using them to quantify the evolution 
of the VDF (see a thorough discussion on the related Schechter-LF issue in MD16A). The Schechter-like
parametrization that we provide must be considered purely a mathematical description, 
not a basis from which to infer physically meaningful parameters.

In order to provide a reference for the very-high-$\sigma$ galaxy population, we have 
estimated the abundance of galaxies with $\sigma > 400~km/s$ (i.e. $\log_{10} \sigma > 2.6$). 
By integrating the VDF, we find a number density of $\sim7.5 \times 10^{-9}$ Mpc$^{-3}$ at $z=0.55$.
This number density implies that just a few
hundred objects with such extreme velocity dispersions may exist within the entire volume of the BOSS CMASS sample
(i.e. $\sim7.8$ Gpc$^{3}$), if the uncertainty in the determination of the VDF is taken into account.

Figure~\ref{fig:vdf_boss} shows a slow evolution in the RS VDF within the narrow redshift range considered.
Such evolution, which makes the RS VDF shift slightly towards higher $\sigma$ as we move to higher redshift
(while the overall normalization remains fairly constant), is not statistically 
significant, given the uncertainties in the determination of the VDF. It is worth noting, 
however, that the result that high-$\sigma$ galaxies are progressively more abundant at higher redshift, if confirmed, 
would be inconsistent with previous findings from \citealt{Chae2010}, 
who predicted the opposite trend.

The L-$\sigma$ relation used to compute the RS VDF shown in Figure~\ref{fig:vdf_boss} corresponds to 
the velocity dispersion averaged within the effective radius, $R_e$. In MD16B,
we perform an {\it{aperture correction}} (AC) in order to correct for the fact that the angular size of the BOSS fibers (radius $R_{\mathrm{aperture}} = 1$ arcsec) 
probes progressively larger physical scales as we move to higher redshift in the sample. 
The AC is also necessary because it facilitates comparison between independent results. In 
MD16B, the following expression is used to relate the observed velocity dispersion, $\sigma_{obs}$, and 
the velocity dispersion averaged within $R_e$, $\sigma(<R_e)$:

\begin{equation}
\sigma_{obs} / \sigma(<R_e) = 0.98~(R_e/R_{\mathrm{aperture}})^{0.048} 
\label{eq:aperture}
\end{equation}

Note that the $<\log_{10} \sigma>$ - apparent magnitude relation is aperture-corrected a posteriori
in MD16B using Equation~\ref{eq:aperture} and the $<\log_{10} R_{e}>$ - apparent magnitude
relation. It is shown in MD16B that the effect on the zero-point of the L-$\sigma$ relation is very small (see
Equation 26 in MD16B). The AC has, conversely, a significant effect on the slope of the L-$\sigma$ relation
in BOSS. The correction on the slope is equal to the slope of the $<\log_{10} R_{e}>$ - magnitude relation 
times the exponent $0.048$ (see Equation 27 in MD16B). As shown in Figure 2 from MD16B, 
the $<\log_{10} R_{e}>$ - magnitude relation is relatively steep, which implies a non-negligible correction. This effect has an obvious 
physical interpretation: brighter, and hence larger, galaxies must have their velocity dispersions corrected 
by a different factor than fainter, and hence smaller, galaxies, for a given fixed angular aperture.

Since there is an unavoidable level of uncertainty associated with the AC, it is useful 
to show the raw RS VDF \footnote{Note, also, that certain applications
of the VDF may require the raw VDF instead of the aperture-corrected VDF.}. This function is presented and compared with 
the aperture-corrected RS VDF in Figure~\ref{fig:vdf_nonac}. Not surprisingly, given that the
aperture correction increases the slope of the L-$\sigma$ relation by $40\%$,
the difference between these two functions is significant, with the aperture-corrected VDF being above the raw 
VDF for the majority of the $\sigma$ range considered. Note that, as mentioned above, the shape
of the VDF is more sensitive to changes in the L-$\sigma$ relation than it is to 
variations in the LF.

Finally, since the modified Schechter fit does not provide a perfect description of the 
high-mass RS VDF, in Table~\ref{tb:vdf_055} we provide the data points for the $z=0.55$
VDF (aperture-corrected and raw) within the range of interest. 

\begin{table}
\centering
\begin{tabular}{ c | c | c }
\hline
\hline
$V = \log_{10} \sigma$ & $\log_{10} \Phi(V)$ (AC) &  $\log_{10} \Phi(V)$ (Raw) \\
\hline
\hline
$2.35$ &      $ -2.137 \pm 0.084 $ &      $ -2.407 \pm 0.082$ \\
\hline
$2.37$ &      $ -2.232 \pm 0.081 $ &      $ -2.509 \pm 0.079$ \\
\hline
$2.39$ &      $ -2.351 \pm 0.076 $ &      $ -2.632 \pm 0.074$ \\
\hline
$2.41$ &      $ -2.499 \pm 0.070 $ &      $ -2.780 \pm 0.069$ \\
\hline
$2.43$ &      $ -2.681 \pm 0.062 $ &      $ -2.955 \pm 0.063$ \\
\hline
$2.45$ &      $ -2.899 \pm 0.056 $ &      $ -3.160 \pm 0.059$ \\
\hline
$2.47$ &      $ -3.157 \pm 0.057 $ &      $ -3.399 \pm 0.060$ \\
\hline
$2.49$ &      $ -3.458 \pm 0.069 $ &      $ -3.673 \pm 0.069$ \\
\hline
$2.51$ &      $ -3.806 \pm 0.093 $ &      $ -3.986 \pm 0.087$ \\
\hline
$2.53$ &      $ -4.202 \pm 0.126 $ &      $ -4.339 \pm 0.113$ \\
\hline
$2.55$ &      $ -4.650 \pm 0.168 $ &      $ -4.737 \pm 0.147$ \\
\hline
$2.57$ &      $ -5.150 \pm 0.217 $ &      $ -5.180 \pm 0.187$  \\
\hline
$2.59$ &      $ -5.706 \pm 0.272 $ &      $ -5.670 \pm 0.233$ \\
\hline
$2.61$ &      $ -6.319 \pm 0.335 $ &      $ -6.210 \pm 0.285$ \\
\hline
$2.63$ &      $ -6.990 \pm 0.404 $ &      $ -6.801 \pm 0.343$ \\
\hline
$2.65$ &      $ -7.720 \pm 0.480 $ &      $ -7.445 \pm 0.408$ \\
\hline
\hline
\end{tabular}
\caption{Data points at the high-mass end for both the aperture-corrected and 
the raw RS VDF at $z=0.55$.}
\label{tb:vdf_055}
\end{table}

\begin{figure*}
\begin{center}
\includegraphics[scale=0.45]{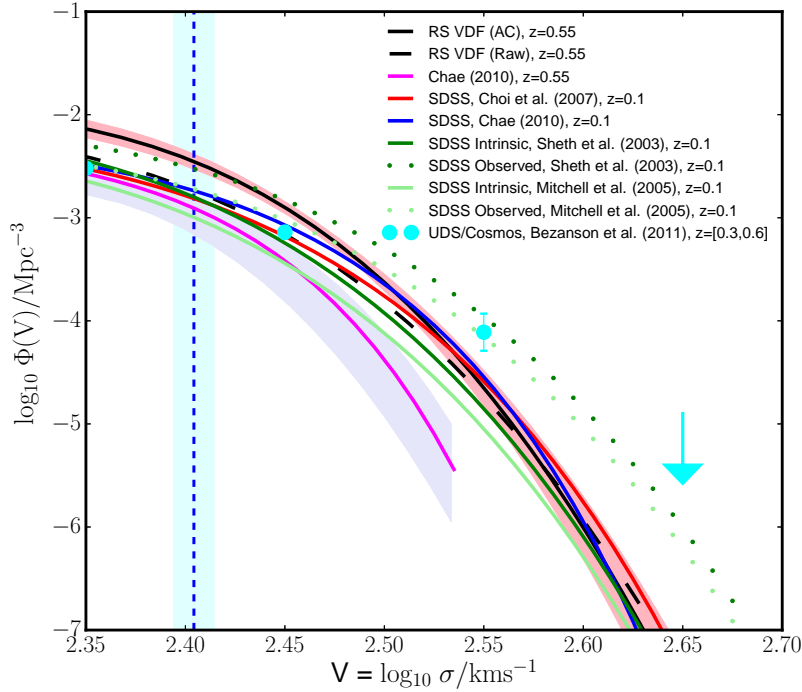}
\caption{Comparison of the high-mass RS VDF at $z=0.55$ (both aperture-corrected and raw) with previous results. All 
$z\sim0.1$ SDSS-I results are direct spectroscopic measurements. The $z\sim0.5$ results from  
Bezanson et al. (2011) and Chae (2010) are indirect measurements based on 
photometrically-derived predictions and strong-lensing statistics, respectively. 
The Chae (2010) $z=0.55$ VDF has been obtained by evaluating at the aforementioned redshift the parametric evolutionary model 
provided by the author. The vertical line indicates the $50\%$ completeness limit of the sample
in velocity dispersion, $\sigma_{50}$; the shaded region around this line
shows the range between completeness values of $2\%$ and $98\%$, respectively. Shaded regions around the RS VDF indicate the 1-$\sigma$ statistical uncertainty.
Shaded regions around the Chae (2010) $z=0.55$ VDF correspond to the $68\%$ confidence limit quoted by the author.
} 
\label{fig:vdf_all}
\end{center}
\end{figure*}

\subsection{The evolution of the VDF from $z\sim0.55$ to $z\sim0$} 
\label{sec:evolution}

In Figure~\ref{fig:vdf_all}, we compare the RS VDF at $z=0.55$ with published direct  
measurements of the low-redshift VDF (some of which are based on the same data set as one another) and 
predicted/inferred VDFs at $z=0.55$, at the high-mass end. With regard to 
$z=0.1$ results, we show direct measurements from \cite{Sheth2003} (both the observed and the intrinsic VDF), 
\cite{Mitchell2005} (both the observed and the intrinsic VDF), \cite{Choi2007} and \cite{Chae2010}\footnote{The 
observed VDFs are the result of convolving the intrinsic VDFs with measurement errors.}.
All these measurements use SDSS data, although they differ in the particular sample, selection or method employed. 
\cite{Sheth2003} use the sample of $\sim5000$ ETGs selected by \cite{Bernardi2003a} from the SDSS Early Data Release. 
The \cite{Mitchell2005} estimate is essentially an update on this measurement using the subsequent 30,000-ETG sample 
of \cite{Bernardi2005}, where also some minor changes in the ETG selection are implemented. \cite{Choi2007} develop their 
own ETG selection, to be applied to a much bigger SDSS parent sample of more than 300,000 galaxies. Finally, 
the measurement from \cite{Chae2010} is a modification of the VDF from \cite{Choi2007} that employs a Monte Carlo
method to correct for low-$\sigma$ incompleteness. In the remainder of this section, we address the comparison 
with previous low-z and $z=0.55$ results separately.

\subsubsection{Previous direct measurements at $z\sim0$}

The first thing to notice from Figure~\ref{fig:vdf_all} is the significant scatter found
among different measurements of the low-redshift (intrinsic) VDF at the high-$\sigma$ end. This is not 
surprising, since the SDSS covers mostly the low-to-intermediate mass range and the measurements
are dominated by these ranges. Small differences in the sample selection or method employed will be especially noticeable 
at the high-mass end. Within the scatter, our $z=0.55$ RS VDF and the low-z intrinsic VDFs converge to 
similar number densities at the very high-mass end, i.e., $\log_{10} \sigma \gtrsim 2.47$. At lower velocity dispersions, however,
the $z=0.55$ RS VDF progressively separates from the low-z estimates, so that 
the number density of RS galaxies with $\log_{10} \sigma \sim 2.40 - 2.45$ (i.e., $\sim \log_{10} \sigma_{50}$) is higher at $z=0.55$.

The comparison between the $z=0.55$ RS VDF and the $z=0.1$ ETG VDFs is subject to the fact that 
the samples have been selected according to different criteria. The low-z results are obtained using  
a ``traditional" morphologically-selected ETG sample (see \citealt{Bernardi2003a, Bernardi2005} 
and \citealt{Choi2007} for details). Our RS was identified from a photometric-deconvolution 
procedure, completely based on the phenomenology of the colour-colour plane. 
Despite the reported discrepancies, the convergence found at high velocity dispersions is reassuring, 
given the different approaches taken to define the low-z ETG population and the high-z RS population, respectively.

\begin{figure}
\begin{center}
\includegraphics[scale=0.35]{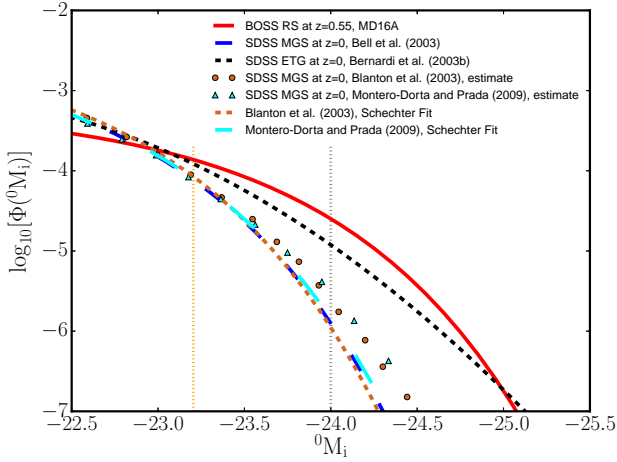}
\caption{The LF for the RS at $z=0.55$ is compared 
with several low-z SDSS LF estimates, in the rest-frame i band. We include the LF 
for the SDSS Main Galaxy Sample (MGS) Data Release 2 (DR2), the SDSS MGS DR6, 
the Early Data Release (EDR) and the ETG sample. The vertical lines 
show, from left to right, the BOSS $50 \%$ - completeness faint limit and the approximate SDSS bright limit, respectively.}
\label{fig:lf_compare}
\end{center}
\end{figure}

Independently of methodological and selection differences, the $z=0.55$ and $z=0$ VDF disagreement at ``lower" velocity dispersions is not surprising 
in consideration of the quantitative differences between the elements from which the VDFs at high and low-z are computed.
In order to illustrate this, we start by focusing on the LF. In Figure~\ref{fig:lf_compare},  we compare the RS LF at $z=0.55$ with the SDSS Main Galaxy Sample (MGS) LF of 
\cite{Blanton2003} and \cite{MonteroDorta2009} and with the SDSS ETG LF of \cite{Bernardi2003a}. In order to facilitate
comparison, we show rest-frame LFs, i.e., K-corrected to $z=0$ (for the MGS LFs we subtracted a factor 
$2.5 \log_{10} (1+z_0)$, where $z_0 = 0.1$, to approximately account for the fact that these LFs where K-corrected to $z=0.1$). 
The dotted vertical line provides a reference for the completeness bright limit of the SDSS LFs. By integrating 
the \cite{Blanton2003} LF,  we find that the total number of galaxies with $^{0}M_i < - 24$ in the SDSS Data Release 2 is no more than 
a few dozens. A similar ad hoc bright limit is likely to be fainter for the \cite{Bernardi2003a} LF, since 
an additional ETG selection is applied (note also that the ETG LF that we show is based on an earlier version of the SDSS data set). 
Figure~\ref{fig:lf_compare} shows that the $z=0.55$ RS LF is significantly brighter than the $z=0$ LF, an effect that is due 
to evolution ($\sim 0.7$ mag). It is interesting that the ETG SDSS LF and the combined SDSS LFs differ considerably at the high-mass end, where
an agreement would be expected. This discrepancy should, nevertheless, be interpreted with caution, in light 
of the significant level of extrapolation of fits from lower luminosities that these low-z LFs are subject to at these number densities.

\begin{figure}
\begin{center}
\includegraphics[scale=0.5]{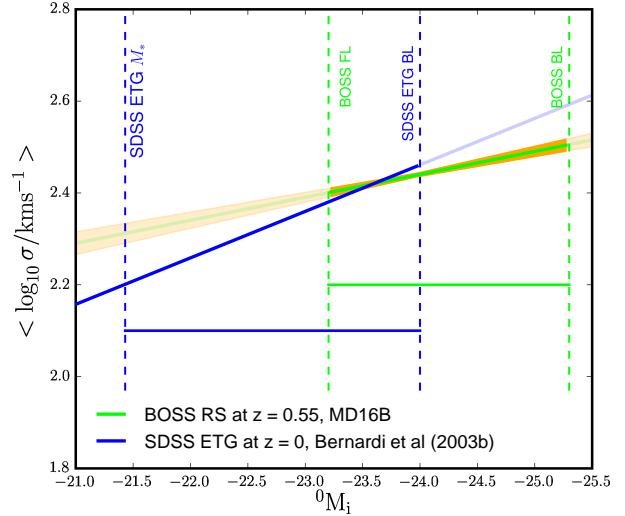}
\caption{The $<\log_{10} \sigma>$ - absolute magnitude relation in the rest-frame i band 
from MD16A at $z=0.55$ and from Bernardi et al. (2003b) at $z=0$.  The shaded regions 
represent the uncertainty in the determination of the former. The vertical lines 
show, from left to right: the characteristic absolute magnitude in the ETG sample, 
the BOSS $50 \%$ - completeness faint limit (FL), the approximate ETG-sample bright limit (BL) and the
BOSS FL.}
\label{fig:lsigma_compare}
\end{center}
\end{figure}

 As a result of the different magnitude/mass ranges covered by 
the SDSS and BOSS, the L-$\sigma$ relation measured from these two data sets at the 
high-mass end is also significantly different. The L-$\sigma$ relation measured 
from low-z ETG samples approximates the canonical Faber-Jackson form $L \propto \sigma^{4}$ 
at the high-mass end, since the measurement is dominated by lower-mass ranges. In Figure~\ref{fig:lsigma_compare}
we show the $<\log_{10} \sigma>$ - absolute magnitude relation from MD16B and 
\cite{Bernardi2003b}, in the rest-frame i band. Figure~\ref{fig:lsigma_compare} is intended to illustrate the 
effective magnitude ranges where each measurement is valid (darker colours) and 
the ranges where each measurement must be considered pure extrapolation (lighter colours).  
Figure~\ref{fig:lsigma_compare} also shows the approximate range where the L-$\sigma$ relation breaks, i.e., $^{0}M_i \simeq -23.5$. 
Importantly, the extrapolation of a high-mass relation towards low-mass ranges in BOSS
has little effect on the high-mass end\footnote{For the sake of completeness, it is noteworthy that the high-mass curvature of the L-$\sigma$ relation 
was detected in the SDSS-I in subsequent works (\citealt{Desroches2007, HydeBernardi2009, Bernardi2011}),
but a definite quantification of the high-mass slope from the SDSS-I never emerged.}. On the 
other hand, the VDF is also sensitive to the scatter measured for the L-$\sigma$ relation. In this sense, 
although the \cite{Sheth2003} results account for the dependence of the scatter as a function of luminosity, our implicit 
$L - \sigma$ relation has, overall, a smaller scatter (of only $0.047$ dex in $\log_{10} \sigma$) than what have been measured at low-z.

Finally, it is also noteworthy that the $z=0.55$ and $z=0$ VDF disagreement at ``lower" velocity dispersions is of similar 
magnitude as the difference between the AC $z=0.55$ VDF and the raw $z=0.55$ VDF (see black dashed line in Figure~\ref{fig:vdf_all}).
We cannot discard the possibility that uncertainties in the determination of this correction in both samples could 
be responsible for the differences found.

\subsubsection{Previous indirect measurements at $z\sim0.55$}

At $z\sim0.55$, direct spectroscopic measurements of the VDF have not been reported prior to this work. We have over-plotted in 
Figure~\ref{fig:vdf_all} the indirectly-derived VDFs from \cite{Bezanson2011}, within the redshift range $0.3<z<0.6$,
and \cite{Chae2010}, at $z=0.55$. The \cite{Bezanson2011} VDF is a prediction based on photometric properties, which were calibrated to 
low-redshift dynamical relations. For the \cite{Chae2010} inference, which is based on strong-lensing statistics, we show 
the evolutionary model provided by the authors for the modified-Schechter function, evaluated at $z=0.55$. The uncertainty 
associated to this inference, represented by a shaded region in Figure~\ref{fig:vdf_all}, is assumed here 
to be of the same magnitude as the one quoted by \cite{Chae2010} at $z=1$ ($68\%$ confidence limit in Figure 3 from 
\citealt{Chae2010}).

The \cite{Bezanson2011} VDF sits well within the low-z scatter at $\log_{10} \sigma \lesssim 2.5$,
but predicts much larger number densities than the low-z VDF and our $z=0.55$ results 
at higher $\sigma$ (note, however, that the highest-$\sigma$ data point is an upper limit). 
In \cite{Bezanson2012}, an explanation for this high-$\sigma$ upturn is provided. By repeating the measurement splitting galaxies into star-forming and quiescent,
the authors show that the scatter between measured and inferred velocity dispersion, when combined with the steep high-$\sigma$ 
end of the VDF, can produce the same observed upturn that causes the aforementioned tension with low-z results. 
This result suggests that the upturn found by \cite{Bezanson2011} is caused by an
observational effect that, if understood, could be deconvolved from their measurement.

The inferred VDF from \cite{Chae2010} is consistent with the low-z VDF at lower velocity dispersions, but
suggests the opposite type of evolution than the one reported by \cite{Bezanson2011} at the high-$\sigma$ end, where the \cite{Chae2010} VDF number densities are 
significantly smaller than the low-z results (note that this VDF is only provided for $\log_{10} \lesssim 2.53$). 
The uncertainties quoted for these results are large enough, however, that a no-evolution scenario for high-$\sigma$ galaxies
cannot be completely discarded. 

The main conclusion to be drawn from this comparison is that the huge scatter found between previous indirect measurements/predictions and the large errors associated to these 
analyses at $z=0.55$ illustrates the significance of the first direct spectroscopic measurement that we report in this work.

\begin{table*}
\centering
\begin{tabular}{ c | c | c | c | c | c}
\hline
\hline
Source & Redshift & $\Phi_* (\times 10^{-3} \rm{Mpc}^{-3})$ & $\sigma_*$ (km/s)& $\alpha$ & $\beta$ \\
\hline
BOSS RS VDF, AC (this work) &$0.55$& $6.97\pm1.06$ &$118.86\pm12.40$ & $6.75\pm0.99$ & $2.37\pm0.14$ \\
\hline
BOSS RS VDF, Raw (this work) &$0.55$& $5.48\pm1.52$ & $116.40\pm15.12$ & $5.03\pm1.03$ & $2.23\pm0.17$ \\
\hline
\citet{Chae2010} & $0.55$ & $3.16^{+1.84}_{-1.36}$ & $216.31^{+14.38}_{-26.02}$ & $1.05^{+0.32}_{-0.29}$ & $4.58^{+1.39}_{-1.28}$ \\
\hline
\citet{Choi2007} & $0.1$ & $2.74$ & $161.2\pm5$ & $2.32\pm0.1$ & $2.67\pm0.07$ \\
\hline
\citet{Chae2010} & $0.1$ & $4.46$ & $217.0$ & $0.85$ & $3.72$ \\
\hline
\citet{Mitchell2005}, Intrinsic & $0.1$ & $1.4\pm0.1$ & $88.8\pm17.7$ & $6.5\pm1.0$ & $1.93\pm0.22$ \\
\hline
\citet{Mitchell2005}, Observed & $0.1$ & $1.4\pm0.1$ & $88.8\pm17.7$ & $6.5\pm1.0$ & $1.8\pm0.22$ \\
\hline
\citet{Sheth2003}, Intrinsic & $0.1$ & $2.2\pm0.1$ & $88.8\pm17.7$ & $6.5\pm1.0$ & $1.93\pm0.22$ \\
\hline
\citet{Sheth2003}, Observed & $0.1$ & $2.2\pm0.1$ & $88.8\pm17.7$ & $6.5\pm1.0$ & $1.8\pm0.22$ \\
\hline
\hline
\end{tabular}
\caption{Best-fit parameters of the modified Schechter function (Equation 11) for the high-mass VDF 
at $z=0.55$ and for several independent measurements of the low-redshift VDF. The 
Chae (2010) $z=0.55$ VDF parameters have been obtained by evaluating at the aforementioned redshift the parametric evolutionary model 
provided by the author. The best-fit modified-Schechter functions provide a good description of the VDF, but slightly underestimate it in 
the very-high-$\sigma$ range. Also, it is important to bear in mind that Schechter-like 
parameters are not adequate to quantify the evolution of the high-mass end of the VDF. The best-fit
Schechter function provided in this work for the BOSS VDF must be considered purely a mathematical description, 
 not a basis from which to infer physically meaningful parameters.}
\label{tb:vdf_fits}
\end{table*}

In Table~\ref{tb:vdf_fits}, we provide a comparison of best-fit modified-Schechter parameters for the
different measurements discussed in this Section, both at low and high redshift. The large scatter in the best-fit parameters shown in Table~\ref{tb:vdf_fits} confirms 
the idea that the Schechter-like parameters are highly covariant and nearly degenerate at the high-mass end, so a comparison based of their values would 
be misleading. Again, in the context of this work, we consider this Schechter-like function a mathematical fitting function, 
not a basis from which to infer physically meaningful parameters.

\subsection{The Sample PDF} 
\label{sec:pdf}

The sample PDF, $P_{CMASS}^{[RS]}$, provides the probability density per unit 
stellar velocity dispersion $\sigma$ of finding an RS galaxy at a given redshift within the sample. 
$P_{CMASS}^{[RS]}$ is therefore uncorrected for incompleteness. This sample PDF is
of interest because it is directly relevant to statistical studies of gravitational lensing within
the BOSS sample \citep{Arneson2012}. $P_{CMASS}^{[RS]}$ can be obtained for a given 
redshift by computing the following expression:

\begin{eqnarray} \displaystyle
P_{CMASS}^{[RS]}(\sigma) = \frac{\int{ dL \phi(L)~p(\sigma|L) C(L)}}{\int\int{ d\sigma dL \phi(L)~p(\sigma|L) C(L)}}
\label{eq:pdf}
\end{eqnarray}

\noindent which takes into account not only the LF, $\phi(L)$, but also 
the completeness in the sample, $C(L)$. $C(L)$ is defined as the fraction of galaxies with luminosity $L$ in intrinsic space
that is expected to pass the CMASS selection criteria, taking into account the photometric 
uncertainties present in the sample. $C(L)$ is provided in MD06A\footnote{Note that 
the CMASS sample is selected in colour-magnitude space, so the photometric selection 
function had to be converted into the one-dimensional completeness function $C(L)$, for a given redshift.}.

\begin{figure}
\begin{center}
\includegraphics[scale=0.35]{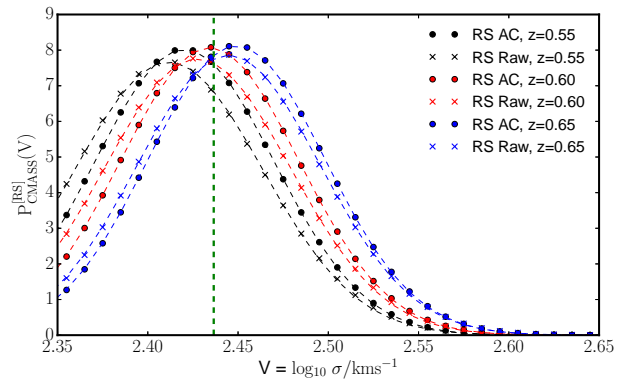}
\caption{Aperture-corrected (AC) and raw sample PDF for the high-mass RS at $z=0.55,0.60$ and $0.65$. Dashed lines
show best-fit Gaussian functions (see text). The vertical line
shows the approximate $50\%$-$\sigma$-completeness threshold in the sample, $\sigma_{50}$.} 
\label{fig:vdf_pdf}
\end{center}
\end{figure}

In Figure \ref{fig:vdf_pdf}, we show the sample PDF at redshifts $z=0.55,0.60$ and $0.65$, obtained from both the raw
and the aperture-corrected L-$\sigma$ relation in each case. Figure \ref{fig:vdf_pdf} shows that there is a higher
probability of finding high-$\sigma$ galaxies at higher redshifts in the sample. The sample
PDF, although not required to be Gaussian, is extremely well characterized by a Gaussian function 
in $\log_{10} \sigma$ of the form:

\begin{equation}
P_{CMASS}^{[RS]} (\log_{10} \sigma) = \frac{1}{\sqrt{2 \pi} s_{PDF}} \exp\left[\frac{-(\log_{10} \sigma - \mu_{PDF})^2}{2 s_{PDF}^2}\right]
\label{eq:gaussian}
\end{equation}

Best-fit parameters for both the aperture-corrected and the raw sample PDF 
at redshifts $z=0.55,0.60$ and $0.65$ are listed in Table~\ref{tb:sample_pdf}. The vertical line in Figure \ref{fig:vdf_pdf}
indicates the approximate $50\%$-$\sigma$-completeness threshold in the sample, $\sigma_{50}$.
Note that the sample PDF is normalized over all sigma, but care should be exercised when
interpreting the predictions of the sample PDF at low values of sigma, due to incompleteness
in the sample. 

Although this sample PDF calculation ``backs out'' the effects of completeness, it still incorporates
the statistical separation of RS galaxies from BC galaxies to give the sample PDF of the former.
A computation of the combined sample PDF for RS and BC galaxies together
could be made entirely in velocity-dispersion space, in the manner of \cite{Shu2012}, without making use of 
the photometric selection function information from MD16A.

\begin{table}
\centering
\begin{tabular}{ c | c | c | c | c }
\hline
\hline
Type & $\mu_{PDF}$ & $\sigma_{PDF}$ \\
\hline
$z=0.55$, AC & $2.420$ & $0.0497$ \\
$z=0.55$, Raw & $2.412$ & $0.0520$ \\
\hline
$z=0.60$, AC & $2.434$ & $0.0494$ \\
$z=0.60$, Raw & $2.428$ & $0.0514$ \\
\hline
$z=0.65$, AC & $2.449$ & $0.0494$ \\
$z=0.65$, Raw & $2.445$ & $0.0507$ \\
\hline
\hline
\end{tabular}
\caption{Best-fit parameters of the Gaussian model in $\log_{10}\sigma$ for both 
the aperture-corrected (AC) and the raw sample
PDF in 3 different redshifts (see text).}
\label{tb:sample_pdf}
\end{table}

\section{Discussion \& Conclusions}
\label{sec:conclusions}

We provide the first direct spectroscopic measurement of the high-mass end of the VDF at redshift 
$z\sim0.55$ for the RS galaxy population using a spectroscopic sample of $\sim600,000$ massive galaxies taken from 
the SDSS-III BOSS CMASS sample. The $z\sim0.55$ VDF is well described by a modified - Schechter function with best-fit parameters 
$\Phi_* =  (6.97\pm1.06) \times 10^{-3} \rm{Mpc}^{-3}$, $\sigma_* = 118.86\pm12.40$ km/s, $\alpha = 6.75\pm0.99$ and 
$\beta = 2.37\pm0.14$ (this must be considered purely a fitting function, 
not a basis from which to infer physically meaningful parameters). No significant evolution is found within the CMASS restricted redshift range $0.525<z<0.65$. 

Prior to this work, the VDF at higher redshift had only been indirectly inferred from strong-lensing statistics \citep{Chae2010}
or predicted using low-z calibrated photometric information (see \citealt{Bezanson2011}), both methods
being subject to major uncertainties. \cite{Chae2010} used a very small sample of 30 lensing galaxies to infer constrains on the 
$0.3<z<1$ ETG VDF. The photometric predictions (calibrated 
using low-z dynamical relations) from \cite{Bezanson2011} are affected by 
incompleteness in the samples, uncertainties in galaxy-size measurements and 
the validity of the assumption that the relation between the inferred and measured $\sigma$ does not evolve. These independent
indirect measurements are in significant disagreement with each other at the high-mass end, which emphasizes the relevance of our 
direct spectroscopic measurement. They also appear to be in tension with with our 
direct measurement. For \cite{Bezanson2011}, this tension would be alleviated 
if the scatter between measured and inferred velocity dispersions could be deconvolved from their 
measurement, reducing the upturn observed towards higher $\sigma$ \citep{Bezanson2012}.
The inference from \cite{Chae2010} implies very low number densities at $\log_{10} \sigma \sim 2.5$, 
but the large uncertainties quoted by the authors leave some room for agreement.

We have compared our $z = 0.55$ RS VDF with several direct spectroscopic measurements of the $z = 0.1$ ETG VDF.
Despite the scatter found among these previously reported low-z results, the following trend is detected:
the $z = 0.55$ RS VDF is consistent with the $z = 0.1$ ETG VDF at the very high-$\sigma$ end ($\log_{10} \sigma \gtrsim 2.47$) but, 
at lower velocity dispersions, the number density of RS galaxies at $z=0.55$ is higher than the number density of ETG galaxies at $z=0.1$. 
The interpretation of this result as a real evolutionary trend is, however, still subject to differences in the 
way the RS and the ETG galaxy population are identified/selected. If we assume that both selection schemes
are able to identify the same galaxy population at different redshifts, the differences reported 
could be due to the fact that the low-z results at the high-mass end are based on an intermediate-mass
sample (the SDSS-I). In particular, the $L-\sigma$ relation that we measure in MD2016B for the 
$z=0.55$ RS population at the high-mass end, which enters through $p(\sigma|L)$, is significantly steeper and presents a smaller scatter than the one 
measured from these SDSS-I samples (see \citealt{Bernardi2003b, Desroches2007}). 
 
It is worth noting that, at face value, the evolution in the VDF implied by our $z = 0.55$ RS - $z = 0.1$ ETG comparison appears in 
tension with what was predicted/inferred by \cite{Bezanson2011} and \cite{Chae2010}
for a similar redshift range. \cite{Bezanson2011} suggest that the VDF at $0.3<z<0.6$ 
is consistent with the low-z VDF (this is only true towards the high-$\sigma$ end according to our measurement). 
\cite{Chae2010} concludes that the VDF evolves from $z=1$ to $z=0$ in a way that the number density of higher-$\sigma$ 
galaxies is progressively lower (while the number density of intermediate-$\sigma$ galaxies 
remains fairly constant). Note that we detect the opposite trend within our limited CMASS redshift range $0.52<z<0.65$; this
trend is, however, not significant given the uncertainties in our analysis.

On the $z=0.55$ VDF side, we have shown that there is still some
room for improvement, if the uncertainties coming mostly from the L-$\sigma$ relation and 
the aperture correction can be further reduced. This can be achieved in the future by 
incorporating into the analysis data sets obtained using unbiased selection schemes or by 
complementing our results with follow-up higher-S/N spectroscopy of the 
very-high-mass end of the $z=0.55$ galaxy population. In order to place 
tighter constraints on the evolution of the VDF, we plan to extend our analysis to $z=0$
within a consistent forward-modeling framework to identify the intrinsic RS population and compute 
the VDF from BOSS to the SDSS-I. This framework can also be extended to 
higher redshift using upcoming cosmological surveys, such as the 
Extended Baryon Oscillation Spectroscopic Survey (eBOSS, \citealt{Dawson2016}).
Although the general view is that the massive RS population evolves 
in a way that approximates that of a passively evolving galaxy population 
at $z\lesssim1$ (see, e.g., \citealt{Cool2008,Tojeiro2012, Maraston2013, MonteroDorta2016A, MonteroDorta2016B}), a detailed 
characterization has not emerged, and discrepancies with the passive-evolution scenario 
have been reported (see, e.g., \citealt{Bernardi2016}). The importance of a VDF-based 
constraint resides in the fact that is independent 
of SPS models. In addition, the VDF is related to the total mass of galaxies, which provides a different 
angle to the study of massive galaxy evolution, to be combined with the evolution of 
the red SMF (see, e.g., \citealt{Pozzetti2010,Moustakas2013,Bernardi2016}) and the 
red/RS LF (see, e.g., \citealt{Wake2006,Cool2008,Loveday2012,MonteroDorta2016A,Bernardi2016}).

The availability of the VDF for the BOSS sample has potential applications to the
determination of the halo--galaxy connection through the methods of HAM and HOD analysis
applied in terms of velocities rather than the masses of halos and galaxies.
As mentioned above, on the galaxy side of the connection, the VDF eliminates the systematic
uncertainties associated with stellar-mass determination. On the halo side,
peak circular velocity is a more clearly defined quantity than halo mass (see, e.g., \citealt{NFW}).

Finally, we have emphasized the importance of our measurement as a key ingredient for statistical 
computations of the incidence of strong gravitational lensing within the 
BOSS survey. Strong gravitational lenses in BOSS have been identified spectroscopically
in the context of the BOSS Emission-Line Lens Survey (BELLS, see \citealt{Brownstein2012} and \citealt{Shu2016}).

\vspace{9mm} 

We thank Ravi Sheth, Mariangela Bernardi and Rachel Bezanson for providing helpful discussions. 

\vspace{4mm} 

This material is based upon work supported by the U.S. Department of Energy, Office of Science, 
Office of High Energy Physics, under Award Number DE-SC0010331.

The support and resources from the Center for High Performance Computing at the University of Utah are gratefully acknowledged.

Funding for SDSS-III has been provided by the Alfred P. Sloan Foundation, the Participating Institutions, the National Science 
Foundation, and the U.S. Department of Energy Office of Science. The SDSS-III Web site is http://www.sdss3.org/.

SDSS-III is managed by the Astrophysical Research Consortium for the Participating Institutions of the SDSS-III Collaboration 
including the University of Arizona, the Brazilian Participation Group, Brookhaven National Laboratory, University of Cambridge,
University of Florida, the French Participation Group, the German Participation Group, the Instituto de Astrofisica de Canarias, 
the Michigan State/Notre Dame/JINA Participation Group, Johns Hopkins University, Lawrence Berkeley National Laboratory, Max Planck
Institute for Astrophysics, New Mexico State University, New York University, Ohio State University, Pennsylvania State University, University of 
Portsmouth, Princeton University, the Spanish Participation Group, University of Tokyo, The University of Utah, Vanderbilt University, University
of Virginia, University of Washington, and Yale University.

\bibliography{./paper}

\label{lastpage}

\end{document}